\documentclass{moriond}

\def\be{\begin{equation}}
\def\ee{\end{equation}}
\def\bea{\begin{eqnarray}}
\def\eea{\end{eqnarray}}

\newcommand{\Photo}{\includegraphics[height=35mm]{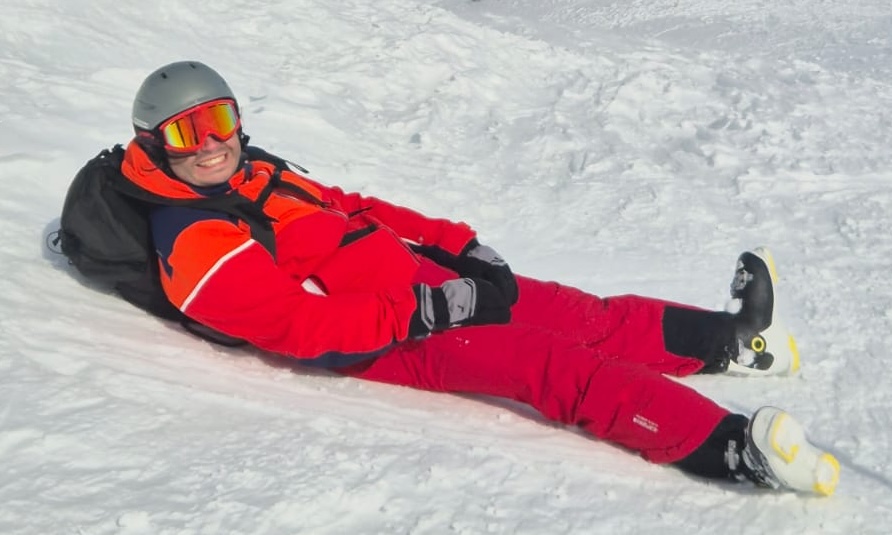}}

\begin{document}
\vspace*{4cm}
\title{Towards the Deployment of the First NectarCAM, a Medium-Sized-Telescope Camera for the Cherenkov Telescope Array Observatory}

\author{ P.~Correa on behalf of the CTAO NectarCAM Collaboration }

\address{Laboratoire de Physique Nucléaire et des Hautes Energies (LPNHE; CNRS/IN2P3)\\4 place Jussieu, F-75252, Paris Cedex 5, France}

\maketitle\abstracts{NectarCAM is a Cherenkov camera designed to detect gamma rays with energies between 80 GeV and 50 TeV. It will equip nine medium-sized telescopes (MSTs) of the Cherenkov Telescope Array Observatory. NectarCAM consists of 1855 pixels distributed over 265 modules. Each pixel consists of a photomultiplier tube that is connected to a NECTAr3 chip. This NECTAr3 chip contains a 12-bit digitizer with a GHz sampling rate, and has a typical readout deadtime of ${\sim}0.7$ $\mu$s. In these proceedings, we highlight the performance of the NectarCAM in terms of time resolution and charge resolution. We also present the latest calibration techniques that were recently implemented for the camera. Finally, we briefly present the current status and plans of the NectarCAM production; the first production-line NectarCAM will be ready for shipment by Summer 2026, and it is planned to equip one of the MST pathfinders of CTAO.}

\section{Introduction}

The Cherenkov Telescope Array Observatory \cite{CTAO_science_2018} (CTAO), currently under construction, is a next-generation instrument targeting the detection of very-high-energy (VHE) gamma rays. Using three types of imaging air-Cherenkov telescopes (IACTs)---large-sized telescopes (LSTs), medium-sized telescopes (MSTs), and small-sized telescopes (SSTs)---CTAO will cover an energy range from a few tens of GeV up to several hundreds of TeV. Within that energy range, CTAO is expected to yield an improvement of a factor $\sim$10 in terms of sensitivity compared to current IACTs. Furthermore, in order to have a complete sky coverage, CTAO will be distributed over two sites: CTAO-North in La Palma (Canary Islands, Spain; 4 LSTs and 9 MSTs in the completed Alpha Configuration) and CTAO-South in Paranal (Atacama Desert, Chile; 14 MSTs and 37 SSTs in the completed Alpha Configuration).

This work focuses on the NectarCAM \cite{Glicenstein_2025}, a Cherenkov camera that will equip 9 MSTs, targeting the intermediate energy range of 0.08--50 TeV and a field of view of ${\sim}8^\circ$. The assembly, integration, and testing of these cameras is performed at CEA/Irfu (Paris-Saclay). Currently, two cameras have been completed: NectarCAM1, a qualification model extensively used for testing and which is to be refurbished, and NectarCAM2, the first production-line camera. Here, we will focus on the performance of NectarCAM1 (Sec.~\ref{sec:performance}) and on the pedestal, gain, and flat-field calibration techniques that were developed using that camera (Sec.~\ref{sec:calibration}). Finally, our plans for the deployment of NectarCAM2 and the future production of the 7 remaining NectarCAMs are presented in Sec.~\ref{sec:outlook}.

\section{Performance of the NectarCAM}
\label{sec:performance}

\begin{figure}
    \centering
    \includegraphics[width=0.4\linewidth]{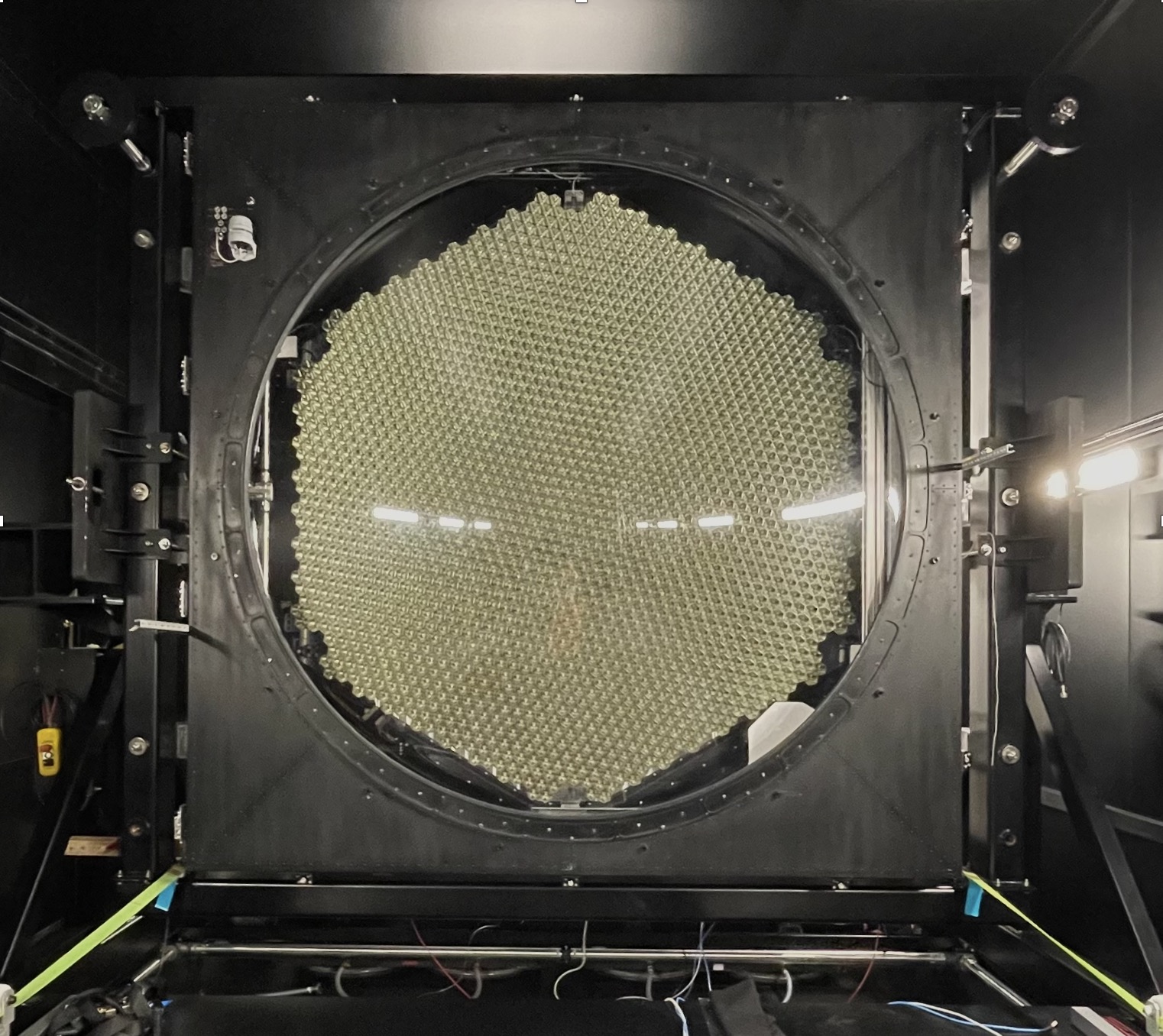}
    \caption{Frontal view of NectarCAM1 in the dark room of the assembly hall at CEA/Irfu. The focal plane of 1855 pixels is visible, as well as the white target (on the bottom right between the focal plane and the shutter).}
    \label{fig:nectarcam}
\end{figure}

The NectarCAM, illustrated in Fig.~\ref{fig:nectarcam}, is constructed following a modular design \cite{Glicenstein_2025}. In particular, 265 focal plane modules (FPMs) constitute the 1855 pixels of the camera. Each FPM consists of 7 pixels, where one pixel is a light guide connected to a photomultiplier tube (PMT). These PMTs are linked to a single front-end board \cite{Bradascio_2024} (FEB) via two gain channels in order to cover the full dynamic range of 0.1--2000 p.e.

After amplification, analog signals of a PMT are processed by its corresponding NECTAr3 chip \cite{Bradascio_2024} on the FEB. The NECTAr3 chip contains a switched-capacitor array with a sampling rate of 1 GHz as well as a 12-bit analog-to-digital converter (ADC). An important feature of the NECTAr3 chip is that it operates in a so-called ``ping-pong" mode, where the analog memory of 1024 ns is split in two. Upon a trigger request, only one of these halves is frozen for the readout and digitization of 60-ns waveforms. As such, the NECTAr3 chip deadtime is reduced to $\sim$0.7 $\mu$s.

\begin{figure}
    \centering
    \includegraphics[width=0.49\linewidth]{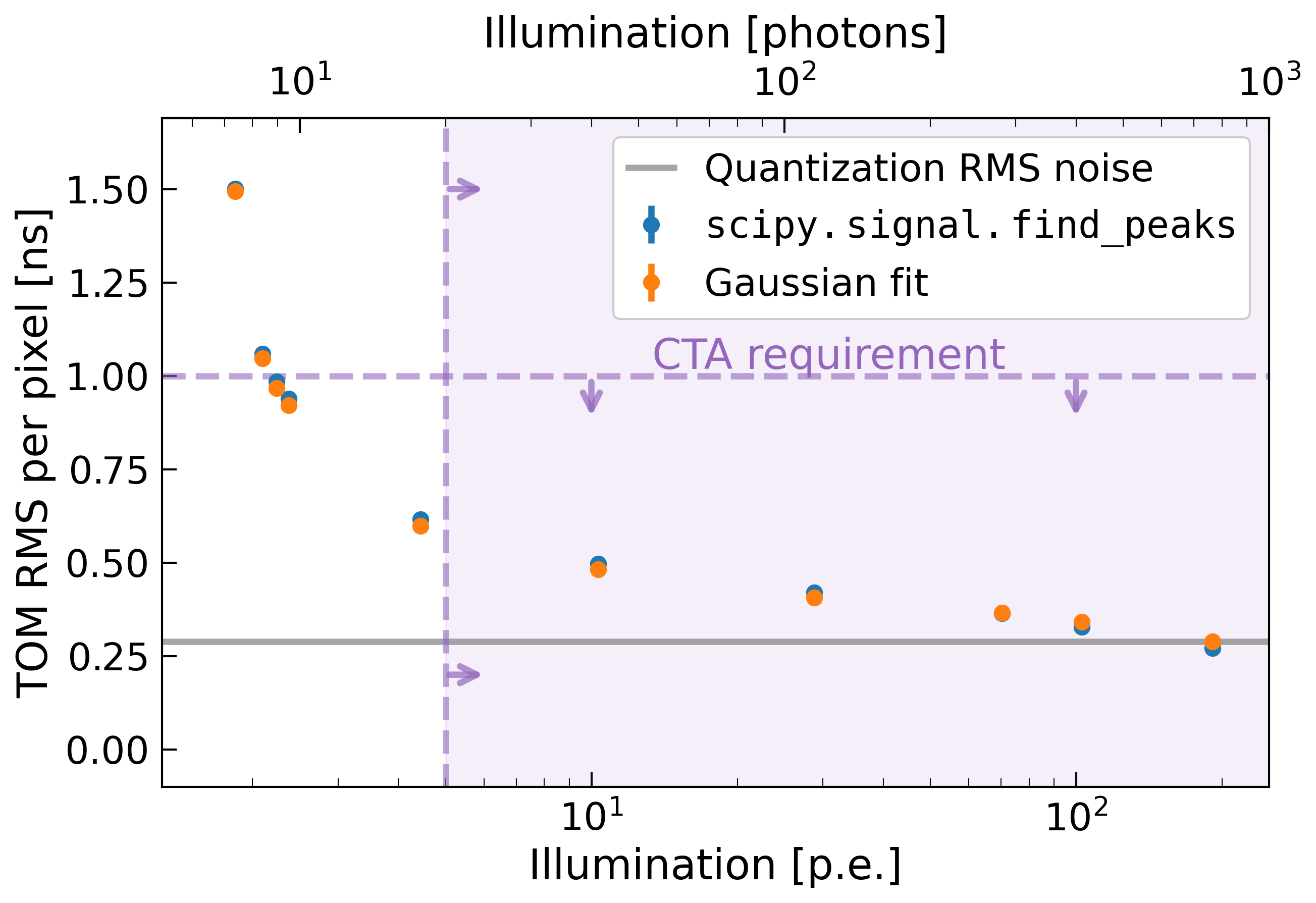}
    \includegraphics[width=0.49\linewidth]{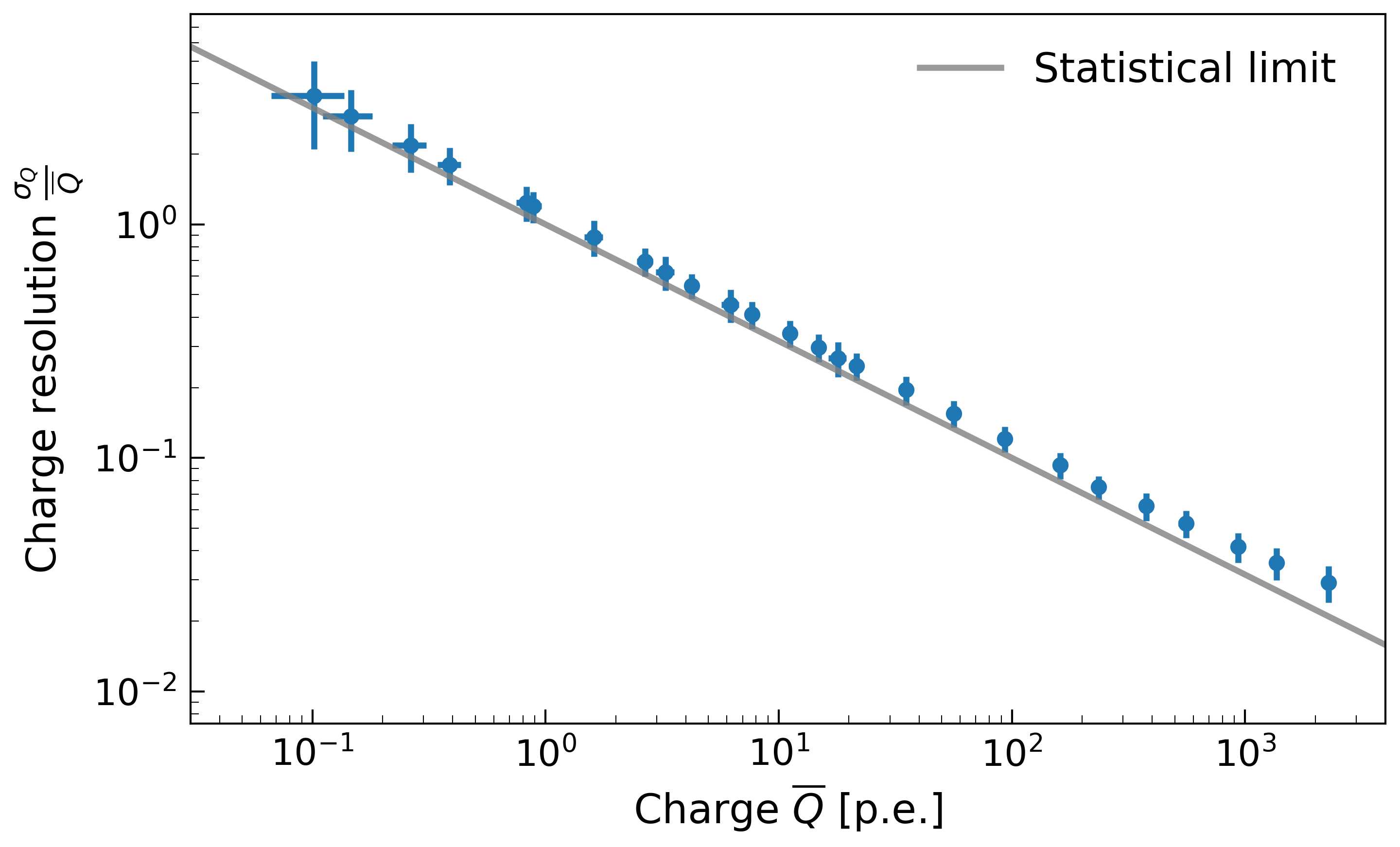}
    \caption{Performance of the NectarCAM in terms of time resolution and charge resolution. \textit{Left}: The RMS of the ToM per pixel as a function of the pixel illumination. The data points show the ToM inferred using two complementary methods. The gray line indicates the pixel timing precision limit. The CTAO requirements are indicated by the shaded region. \textit{Right}: Relative charge resolution as a function of the charge averaged over all pixels. The gray line corresponds to the statistical lower limit described in the text.}
    \label{fig:performance}
\end{figure}

The associated performance of the NectarCAM in terms of time and charge resolution is illustrated in Fig.~\ref{fig:performance}. The time resolution \cite{Bradascio_2023} is characterized by the time of maximum (ToM) of the signal pulse in the recorded waveform. The ToM is computed using two complementary methods. The first method determines the ToM by performing a Gaussian fit to the pulse, while the second method performs an interpolation of the waveform and determines the ToM using the \texttt{scipy.signal.find\_peaks} functionality. As shown on the left panel of Fig.~\ref{fig:performance}, for both methods we achieve a $<$1 ns root mean square (RMS) of the ToM for a camera illumination above 5 p.e., which is well within the CTAO requirements. In addition, the right panel of Fig.~\ref{fig:performance} illustrates that the charge resolution \cite{Bradascio_2024}---where the charge is the integral of the signal pulse (typically with a duration of $\sim$10 ns)---is generally at the statistical lower limit, except at the highest charges where we suffer from PMT saturation effects. This statistical limit arises from Poissonian fluctuations in the number of photoelectrons produced in the PMTs.

\section{Calibration of the NectarCAM}
\label{sec:calibration}

The performance of the NectarCAM discussed in Sec.~\ref{sec:performance} relies on several calibration techniques, which are being developed within the dedicated \texttt{nectarchain} software\cite{Grolleron_2023}. In order to produce calibrated waveforms ($V_{\rm calib}$), raw waveforms recorded by the NECTAr3 chip ($V_{\rm raw}$) are calibrated as
\begin{equation}
    V_{\rm calib}(t) = \frac{ V_{\rm raw}(t) - P(t) }{G} \times F.
\end{equation}
The calibrations of the pedestal $P$, the gain $G$, and flat-field coefficients $F$ are described in the following sections. 


\subsection{Pedestal Calibration}

The first step of the calibration is to subtract the pedestal $P$ (about 250 ADC counts on average) from $V_{\rm raw}$. The pedestal represents the baseline noise due to the night-sky background (NSB) and the camera electronics. At CEA/Irfu, we use a dedicated NSB source to perform the calibration; on-site, this will be achieved by taking images of the night sky. In any case, interleaved pedestal events are recorded by forcing a trigger at a regular rate (e.g.~1 kHz) during the corresponding data-taking run. For each of these runs, we then average out the pedestal waveforms per pixel. Note that this yields an average pedestal waveform $P(t)$ of 60 ns per pixel. This allows us to correct for readout effects in the waveforms, which yield small periodic variations (at a level below 1 ADC count at most) with respect to the waveform baseline. 

\subsection{Gain Calibration}

The second step consists of calibrating the gain $G$, which converts the recorded ADC signal to photoelectrons. The calibration can be performed using a flasher (Sec.~\ref{sec:flatfield}) at either high or low illumination, or using a white target located between the camera shutter and focal plane (Fig.~\ref{fig:nectarcam}). This white target provides a uniform illumination of roughly 40 pixels at a time, and it is moved using a motorized XY table in order to cover the entire camera in roughly 30 minutes.

Two gain-calibration algorithms \cite{Grolleron_2024,Grolleron_2025} are currently being used for NectarCAM. On the one hand, we use the well-established photostatistic method, which infers the gain through the variability in the measured charge over a high-illumination flasher run (Sec.~\ref{sec:flatfield}). On the other hand, we recently implemented a method that directly fits the single photoelectron (SPE) component of the measured charge distribution, for which we use the white target or low-illumination flasher runs. Although the photostatistic method is less computationally expensive, the SPE fit allows for a more precise determination of the gain. In any case, it should be noted that the gain calibration is generally performed for the high-gain data channel (with a nominal gain $G \sim 58~ \rm ADC/p.e.$). In order to cross-calibrate the low-gain channel, we make use of the linearity between the two channels between 7 and 400 p.e., where the high-to-low-gain ratio is 13.1 on average \cite{Bradascio_2024}.

\subsection{Flat-Field Calibration}
\label{sec:flatfield}

The final step of the calibration concerns the flat-field coefficient $F$, which corrects for pixel-to-pixel inhomogeneities in the camera. To do so, we use a dedicated flasher source that illuminates the entire camera \cite{Mikhno_2025}. We recently developed a novel method \cite{Mikhno_2025} that fits a bivariate Gaussian amplitude distribution to the flasher light front instead of modeling it as a simple plane wave. As such, we obtain a description of the light front that is accurate down to the 2\% level. The flat-field coefficient $F$ per pixel is then computed as the ratio between the measured charge in that pixel (averaged over a dedicated flasher run) and its expectation from the Gaussian fit. Note that its nominal value, which corresponds to the flat-field coefficient averaged over the entire camera, is $F = 1$.


\section{Summary and Outlook}
\label{sec:outlook}

In this work, we highlighted the performance of the NectarCAM camera, which will equip 9 MSTs of CTAO. Thanks to its so-called ``ping-pong" mode, the NECTAr3 chip---which lies at the heart of the camera---achieves a deadtime of ${\sim}0.7$ $\mu$s. In addition, the NectarCAM achieves a sub-nanosecond time resolution above 5 p.e., as well as an excellent charge resolution, which is generally at the statistical limit due to Poissonian fluctuations in the PMT photoelectron yield. Furthermore, we highlighted the calibration techniques used for the NectarCAM. In particular, we discussed recent implementations of methods to calibrate the gain and flat-field coefficients of the camera.

Since CTAO is currently ramping up its construction phase, we are set to deliver the 9 NectarCAMs over the coming years. NectarCAM2, the first production-line camera, is currently undergoing its test-readiness review at CEA/Irfu and is due for shipment by Summer 2026. As such, it could equip one of the MST pathfinders currently under construction at CTAO-South, where 2 MSTs and 5 SSTs will be commissioned by 2028. NectarCAM3, which is currently under construction at CEA/Irfu, is planned to be ready for shipment by Summer 2027. Finally, NectarCAM4--9 will be produced at an increased rate after that; NectarCAM1, the qualification model, will be refurbished at the end of the production line. 


\section*{Acknowledgments}

We gratefully acknowledge financial support from the agencies and organizations listed here: \url{https://www.ctao.org/for-scientists/library/acknowledgments/}.

\section*{References}
\bibliography{nectarcam_moriond_2026}


\end{document}